# High-energy threshold reaction rates on 0.8 GeV proton-irradiated thick Pb-target

Yu. E. Titarenko[1,b], V. F. Batyaev[1], A. Yu. Titarenko[1], M. A. Butko[1], K. V. Pavlov[1], R. S. Tikhonov[1], S. N. Florya[1], S. G. Mashnik[2], W. Gudowski[3]

[1] Institute for Theoretical and Experimental Physics (ITEP), 117218 Moscow, Russia
[2] Los Alamos National Laboratory, Los Alamos, NM 87545, USA
[3] Royal Institute of Technology, S - 106 91 Stockholm, Sweden

**Abstract.** This works presents results of activation-aided determination of threshold reaction rates in 92 $^{209}$Bi, $^{nat}$Pb, $^{197}$Au, $^{181}$Ta, $^{169}$Tm, $^{nat}$In, $^{93}$Nb, $^{64}$Zn, $^{65}$Cu, $^{63}$Cu, $^{59}$Co, $^{19}$F, and $^{12}$C samples and in 121 $^{27}$Al samples. All the samples are aligned with the proton beam axis inside and outside the demountable 92-cm thick Pb target of 15-cm diameter assembled of 23 4-cm thick discs. The samples are placed on 12 target disks to reproduce the long axis distribution of protons and neutrons. In June 2006, the target was exposed for 18 hours to a 800-MeV proton beam extracted from the ITEP U-10 accelerator. The proton fluence and the proton beam shape were determined using the $^{27}$Al(p,x)$^{7}$Be monitor reaction. The total number of protons onto the target was $(6.0\pm0.5)\cdot10^{15}$. The reaction rates were determined by the direct gamma-spectrometry techniques. The GC-2518 detectors of 1.8 keV resolution and the DGDK-60V detectors of 2.9 keV resolution at the 1332 keV gamma-line were used to take the measurements. In total, 1196 gamma-spectra have been measured, and about 1500 reaction rates determined. The measured reaction rates were simulated by the MCNPX code using the following databases: ENDF/B6 for neutrons below 20 MeV, MENDL2 for 20-100 MeV neutrons, and MENDL2P for proton cross sections up to 200 MeV. An acceptable agreement of simulations with experimental data has been found.

## 1 Introduction

Many countries all over the world are designing nuclear facilities of new generation where fission is initiated by neutrons produced in interactions of a high current of ~1 GeV proton beam with a heavy target. Spallation Neutron Sources (SNS) and Accelerator Driven Systems (ADS) are examples of this class of facilities. The targets are proposed to be Hg, Pb+Bi, Pb eutectic (liquid targets) or W, Ta (solid targets).

The design of such new generation facilities require new generation of reliable nuclear data. One of the possible fields of data application is activation-based unfolding of high-energy (up to ~1 GeV) neutron spectra inside an ADS target and the near-target zone. The high-energy "tail" in the ADS spectra triggers high-energy-threshold (>10 MeV) reactions, which are not studied in detail in conventional reactor studies. Therefore, excitation functions (EFs) that may be derived from high-energy calculations, or retrieved from available databases require a reliable verification via testing experiments with high-energy proton-irradiated target micromodels.

Our previous work [1] has shown the possibility to build-up such EF's by measuring activation reaction rates (RRs) in samples located on an extended W-Na target induced by 0.8 GeV protons. The objective of the present work is to verify and improve, when needed, the EF's obtained in [1] and to get new EF's based on the latest experiment made under the ISTC Project #2405 [2] using the lead target and induced by 0.8 GeV protons.

## 2 Experimental facility

The Pb target was exposed to the proton beam extracted from the ITEP U-10 synchrotron, which is a ring facility with a 25 MeV energy of proton injection to a ring with the highest proton acceleration energy of 9.3 GeV. Having been accelerated up to a given energy (0.8 GeV in the present experiment), the proton beam consisting of 4 bunches of 250 ns duration each was directed to a transport channel that provided proton extraction with the following parameters: a ~$2\cdot10^{11}$ proton/pulse intensity, an ellipsoidal section with ~10x15 mm axes, and a ~15 pulse/min pulse repetition rate. The Pb target was irradiated for a time T = 64800 s.

The target was assembled of 23 metal discs of 150 mm diameter and 40 mm depth each. The total target length was 920 mm to provide proton beam stoppage. The discs were prepared by casting lead into a graphite lingot with subsequent machining. The task-oriented "rulers" inherent to the disc design were used to irradiate the samples inside the discs. The absolute reaction rates were determined experimentally on $^{209}$Bi, $^{nat}$Pb, $^{197}$Au, $^{181}$Ta, $^{169}$Tm, $^{nat}$In, $^{93}$Nb, $^{65}$Cu, $^{64}$Zn, $^{63}$Cu, $^{59}$Co, $^{27}$Al, $^{19}$F, and $^{12}$C. Fig. 1 shows an overall view of the target with the samples arranged therein. The samples were placed in ether the middle of the side faces of odd discs (in Cd jackets) or task-oriented lead rulers (without Cd jackets) located on the front edges of odd discs.

Prior to irradiation, the target was mounted on a task-oriented table consisting of a cradle and a massive support with micrometering set screws to provide precise adjustment of the cradle with respect to the beam axis.

---

[b] Presenting author, e-mail: Yury.Titarenko@itep.ru

## 3 Proton beam monitoring and normalization

The number of protons on the target and the proton shape were determined using an Al monitor of 170-mm diameter placed at 5 cm before the first disc to intercept the proton beam. Having been irradiated, the monitor was cut into its fragments, most of which were 2x2 cm squares. In each of the fragments, the numbers of the $^7$Be, $^{22}$Na, and $^{24}$Na nuclei produced were determined by the gamma-spectrometry techniques. Since the number of $^7$Be nuclei produced in each of the fragments is proportional to the number of protons through a fragment, the respective results obtained using the $^{27}$Al(p,x)$^7$Be reaction data have made it possible to determine the number of protons through each fragment and, after that, reproduce the proton beam shape.

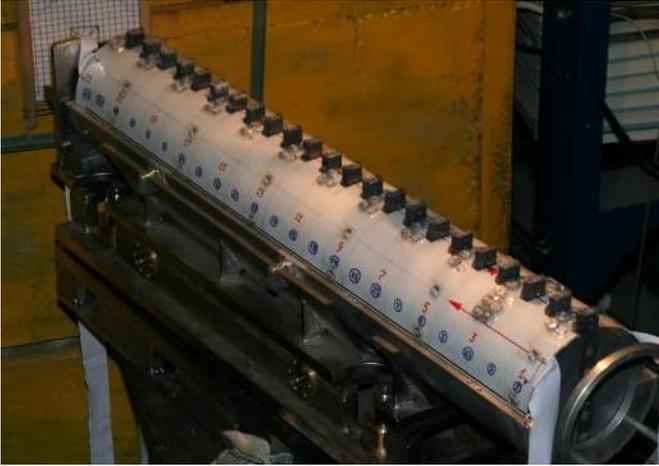

Fig. 1. Overall view of Pb target with experimental samples.

Table 1 presents the results of determining the proton beam parameters, namely, the number ($N$) of particles that hit the target throughout the irradiation time, the positions of Gaussian centers ($x_0, y_0$), their standard deviations ($\sigma_x, \sigma_y$), and the power calculated as $W=N·E/T$, where $E$ is the proton beam energy; $N$ is the total number of protons that hit the target, and $T$ is the total target irradiation time.

**Table 1.** Proton beam parameters.

| Proton number | $(6.0\pm0.5)\cdot10^{15}$ |
|---|---|
| $x_0$ (cm) | $0.27 \pm 0.10$ |
| $\sigma_x$ (cm) | $0.80 \pm 0.06$ |
| $y_0$ (cm) | $0.44 \pm 0.15$ |
| $\sigma_y$ (cm) | $0.70 \pm 0.09$ |
| Power (W) | $11.9 \pm 1.0$ |

## 4 Experimental results

The absolute rates of the independent and cumulative threshold reactions in different experimental samples were determined by the gamma-spectrometry techniques using the expressions [3]

$$R_1^{cum} = \frac{A_1}{N_T \cdot h_2 \cdot e_2 \cdot n_1} \cdot \frac{l_2 - l_1}{l_2} \cdot \frac{1}{F_1}$$

$$R_2^{ind} = \left(\frac{A_2}{F_2} + \frac{A_1}{F_1} \cdot \frac{l_1}{l_2}\right) \cdot \frac{1}{N_T \cdot h_2 \cdot e_2}$$

$$R_2^{cum} = R_2^{ind} + n_1 \cdot R_1^{cum} = \left(\frac{A_1}{F_1} + \frac{A_2}{F_2}\right) \cdot \frac{1}{N_T \cdot h_2 \cdot e_2}$$

where $A_0$, $A_1$, and $A_2$ are factors determined by least-squares fitting the experimental count rates of the parent (*1*) and daughter (*2*) nuclides; $N_T$ is the number of nuclei in an irradiated experimental sample; $\eta_1$ and $\eta_2$ are quantum abundances of the nuclides at energies $E_1$ and $E_2$; $\lambda_1$ and $\lambda_2$ are decay constants; $\varepsilon_1$ and $\varepsilon_1$ are absolute effectivenesses of spectrometer at $E_1$ and $E_2$; $F_1$ and $F_2$ are nuclide saturation functions. The data to be used to identify the nuclides produced have been adopted from the PCNUDAT database [4]. Table 2 lists the measured reaction rates.

**Table 2.** List of the reaction rates measured.

| Samples | Products | Number of reaction rates/ samples |
|---|---|---|
| $^{209}$Bi | $^{203}$Bi, $^{202}$Bi, $^{201}$Bi, $^{203}$Pb(i), $^{203}$Pb, $^{202m}$Pb, $^{201}$Pb, $^{200}$Pb, $^{199}$Pb, $^{198}$Pb, $^{202}$Tl, $^{201}$Tl, $^{200}$Tl(i), $^{200}$Tl, $^{199}$Tl, $^{198}$Tl, $^{197}$Tl, $^{196}$Tl, $^{203}$Hg, $^{192}$Hg, $^{200m}$Au, $^{191}$Pt, $^{172}$Lu, $^{103}$Ru, $^{96}$Tc, $^{99}$Mo, $^{96}$Nb, $^{95}$Nb, $^{95}$Zr, $^{88}$Y, $^{82}$Br, $^{58}$Co. | 327/12 |
| $^{197}$Au | $^{198}$Au, $^{196}$Au, $^{196m}$Au, $^{195}$Au, $^{194}$Au, $^{193}$Au, $^{192}$Au, $^{191}$Au, $^{191}$Pt, $^{189}$Pt, $^{188}$Pt, $^{195m}$Ir, $^{192}$Ir, $^{190}$Ir, $^{188}$Ir, $^{187}$Ir, $^{186}$Ir, $^{186m}$Ir, $^{185}$Ir, $^{184}$Ir, $^{185}$Os, $^{183}$Os, $^{183m}$Os, $^{182}$Os, $^{183}$Re, $^{182}$Re, $^{182m}$Re. | 240/12 |
| $^{181}$Ta | $^{182}$Ta, $^{180}$Ta, $^{178m}$Ta, $^{177}$Ta, $^{176}$Ta, $^{175}$Ta, $^{173}$Ta, $^{181}$Hf, $^{180m}$Hf, $^{175}$Hf, $^{173}$Hf, $^{172}$Hf, $^{171}$Hf, $^{170}$Hf, $^{177}$Lu, $^{172}$Lu, $^{172}$Lu, $^{171}$Lu, $^{171}$Lu, $^{170}$Lu, $^{169}$Lu, $^{169}$Yb, $^{172}$Tm, $^{165}$Tm, $^{154m}$Tb, $^{152m}$Eu. | 157/12 |
| $^{nat}$In | $^{113}$Sn, $^{116m}$In, $^{115m}$In, $^{114m}$In, $^{113m}$In, $^{111}$In, $^{110}$In, $^{109}$In, $^{115}$Cd, $^{112}$Ag, $^{111}$Ag, $^{110m}$Ag, $^{106m}$Ag, $^{105}$Ag, $^{101}$Pd, $^{100}$Pd, $^{105}$Rh, $^{102}$Rh, $^{101m}$Rh, $^{101}$Rh, $^{100}$Rh(i), $^{100}$Rh, $^{99}$Rh, $^{103}$Ru, $^{97}$Ru, $^{96}$Tc, $^{95}$Tc, $^{95m}$Tc, $^{94}$Tc, $^{96}$Nb, $^{90}$Nb, $^{89}$Zr, $^{87m}$Y, $^{87}$Y. | 285/12 |
| $^{169}$Tm | $^{168}$Tm, $^{167}$Tm, $^{166}$Tm, $^{165}$Tm, $^{163}$Tm, $^{161}$Er, $^{160}$Er, $^{157}$Dy, $^{155}$Dy, $^{160}$Tb, $^{156}$Tb, $^{155}$Tb, $^{154}$Tb, $^{153}$Tb, $^{152}$Tb, $^{151}$Tb, $^{150}$Tb, $^{149}$Gd, $^{147}$Gd, $^{156}$Eu, $^{148}$Eu, $^{146}$Eu, $^{145}$Eu, $^{151}$Pm, $^{150}$Pm, $^{148m}$Pm. | 26/1 |
| $^{27}$Al | $^{24}$Na, $^{22}$Na, $^{7}$Be, $^{27}$Mg. | 169/114 |
| $^{59}$Co | $^{57}$Ni, $^{60}$Co, $^{58}$Co, $^{58m}$Co, $^{57}$Co, $^{56}$Co, $^{55}$Co, $^{59}$Fe, $^{52}$Fe, $^{56}$Mn, $^{54}$Mn, $^{52}$Mn, $^{51}$Cr, $^{48}$Cr, $^{48}$V, $^{48}$Sc, $^{47}$Sc, $^{46}$Sc, $^{44m}$Sc, $^{44}$Sc, $^{43}$Sc, $^{47}$Ca, $^{43}$K, $^{42}$K, $^{28}$Mg, $^{24}$Na, $^{22}$Na, $^{7}$Be. | 212/12 |
| $^{93}$Nb | $^{90}$Mo, $^{92m}$Nb, $^{90}$Nb(i), $^{90}$Nb, $^{89}$Zr, $^{88}$Zr, $^{86}$Zr, $^{90m}$Y, $^{88}$Y, $^{80m}$Y, $^{87m}$Y, $^{87}$Y, $^{86m}$Y, $^{86}$Y, $^{85m}$Y, $^{85}$Sr, $^{83}$Sr, $^{86}$Rb, $^{83}$Rb, $^{82}$Br, $^{66}$Ge, $^{57}$Ni, $^{56}$Mn, $^{54}$Mn, $^{52}$Mn, $^{51}$Cr, $^{48}$V, $^{48}$Sc, $^{47}$Sc, $^{46}$Sc, $^{44m}$Sc, $^{44}$Sc, $^{44}$Sc, $^{44}$K | 34/1 |
| $^{64}$Zn | $^{65}$Zn, $^{62}$Zn, $^{64}$Cu, $^{61}$Cu, $^{57}$Ni, $^{58}$Co, $^{58}$Co, $^{58m}$Co, $^{57}$Co, $^{56}$Co, $^{55}$Co, $^{56}$Mn, $^{54}$Mn, $^{52}$Mn, $^{52m}$Mn, $^{51}$Cr, $^{48m}$V, $^{48}$Sc, $^{46}$Sc. | 19/1 |
| $^{65}$Cu | $^{62}$Zn, $^{64}$Cu, $^{61}$Cu, $^{57}$Ni, $^{56}$Ni, $^{60}$Co, $^{58m}$Co, $^{58}$Co, $^{57}$Co, $^{56}$Co, $^{55}$Co, $^{59}$Fe, $^{56}$Mn, $^{54}$Mn, $^{52}$Mn, $^{51}$Cr, | 21/1 |
| | $^{48}$V, $^{48}$Sc, $^{47}$Sc, $^{46}$Sc, $^{44m}$Sc, $^{47}$Ca. | |
| $^{63}$Cu | $^{62}$Zn, $^{64}$Cu, $^{61}$Cu, $^{57}$Ni, $^{56}$Ni, $^{60}$Co, $^{58m}$Co, $^{58}$Co, $^{57}$Co, $^{56}$Co, $^{55}$Co, $^{59}$Fe, $^{56}$Mn, $^{54}$Mn, $^{52}$Mn, $^{51}$Cr, $^{48}$V, $^{48}$Sc, $^{47}$Sc, $^{46}$Sc, $^{44m}$Sc, $^{47}$Ca. | 24/1 |
| $^{59}$Co | $^{57}$Ni, $^{60}$Co, $^{58}$Co, $^{58m}$Co, $^{57}$Co, $^{56}$Co, $^{55}$Co, $^{59}$Fe, $^{52}$Fe, $^{56}$Mn, $^{54}$Mn, $^{52}$Mn, $^{51}$Cr, $^{48}$Cr, $^{48}$V, $^{48}$Sc, $^{47}$Sc-I, $^{47}$Sc, $^{46}$Sc, $^{44m}$Sc, $^{44}$Sc-I, $^{44}$Sc, $^{43}$Sc, $^{47}$Ca, $^{44}$K, $^{43}$K. | 27/1 |
| $^{19}$F $^{12}$C1 | $^{18}$F, $^{11}$C, $^{7}$Be | 3/1 |
| | **TOTAL** | **1544/210** |

Figs. 2-5 show some of the simulated experimental reaction rates. All the measured reaction rates will be presented in the report on ISTC #2405 Project and in the SINBAD nuclear database (OECD Nuclear Energy Agency Data Bank).

## 5 Computational simulation of experimental reaction rates

The measured RRs are integral multiplication of the respective reactions cross sections and particle spectra in the measuring places

$$R_x = R_{n,x} + R_{p,x} = \sum_{i=n,p} \int s_{i,x}(E) \cdot j_i(E) dE .$$

The spectra of particles (protons and neutrons) were simulated with the MCNPX code [5]. The cross sections were obtained as shown in Table 3 using the MENDL [6] (MENDL2p [7] for protons) databases and the LAHET code [8].

**Table 3.** Sources of used here cross sections $s_{i,x}(E)$.

| E (MeV) | Neutrons | Protons | |
|---|---|---|---|
| < 100 | MENDL2 | MENDL2p* | + EXFOR |
| > 100 | LAHET | LAHET | |

* - up to 200 MeV,

The input cross section data were analyzed with the view of their internal consistency and accord with the experimental data available from the EXFOR database. Our analysis of the RRs calculated from the input cross sections shows that some of the input cross sections should be corrected. The analysis and correction procedures are described in detail in [9], where the corrections are shown to permit satisfactory agreements with the experimental RRs.

The results of computational simulating the reaction rates are shown in Figs. 2-5 together with experimental data. The figures show satisfactory agreement between simulated and experimental values for most of the reaction rates. However, there are a number of systematic deviations of simulations from the data, namely,
- The reaction rates simulated for the axis points on first disks are underestimated by factors of 1.5 – 3.0 as compared with experimental data, because the real lateral distribution of the proton beam may differ from the Gaussian, used in simulations (see [9]);
- as a rule, the reactions that show a high production threshold (~100 MeV and higher) are underestimated by a factor of up to 3 and even higher at the surface of the first disk. This can be related to high energy proton presence on the surface of the first disk, i.e. the proton beam lateral distribution is much broader than it is in case of a Gaussian shape.

Other calculations-experiment differences are not systematic and can be explained by the fact that the cross section database used here are insufficiently accurate.

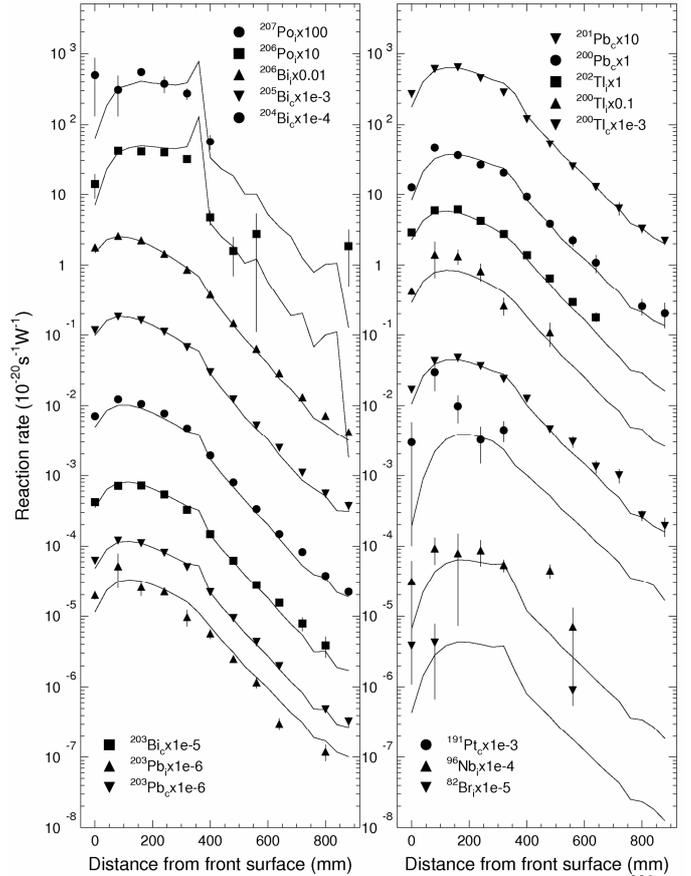

Fig. 2. Experimental and calculated distributions of the $^{209}$Bi reaction products along the surface outside the Pb target.

## 6 Conclusion

The results presented indicate that, with rare exclusions, the EFs obtained in [1] provide good agreement with new measured RRs. Further studies will be aimed at comparison of simulations with full amount of reaction rates measured. A special attention will be paid to simulate properly the proton beam shape.

## 7 Acknowledgement

This work has been carried out under ISTC#2405 Project [2] supported by the USA. The work has been also supported by the Federal Atomic Energy Agency of Russia and, in part, by the U. S. Department of Energy at Los Alamos National Laboratory under Contract DE-AC52-06NA25396.

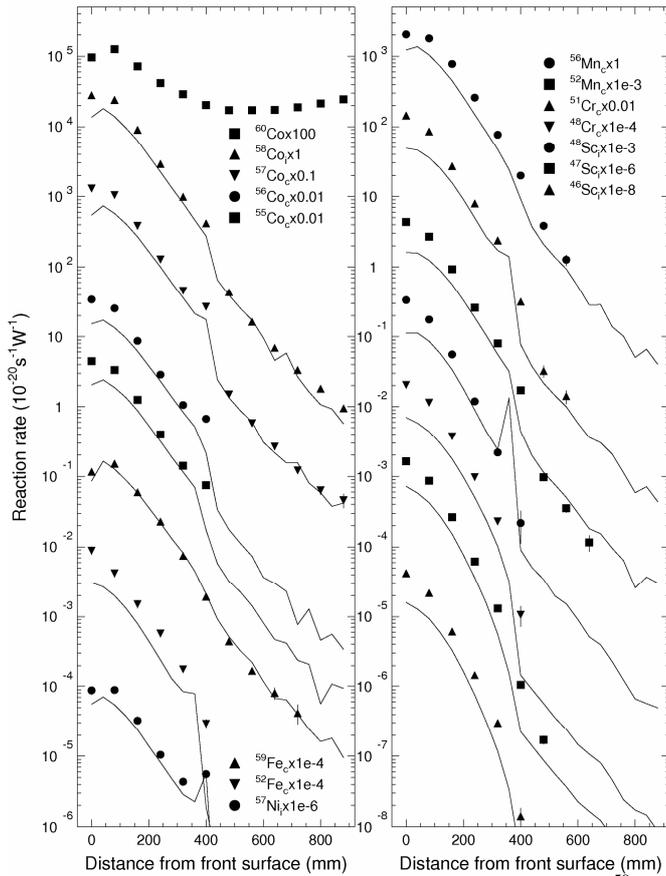

Fig. 3. Experimental and calculated distributions of the $^{59}$Co reaction products along the axis of the Pb target.

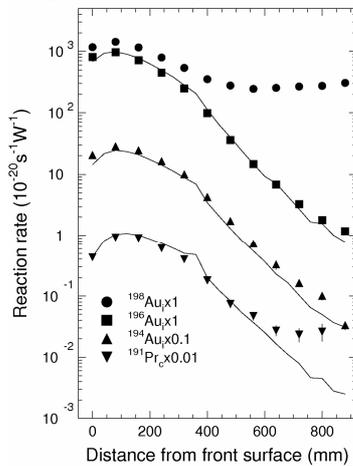

Fig. 4. Experimental and calculated distributions of the $^{197}$Au reaction products along the surface of the Pb target.

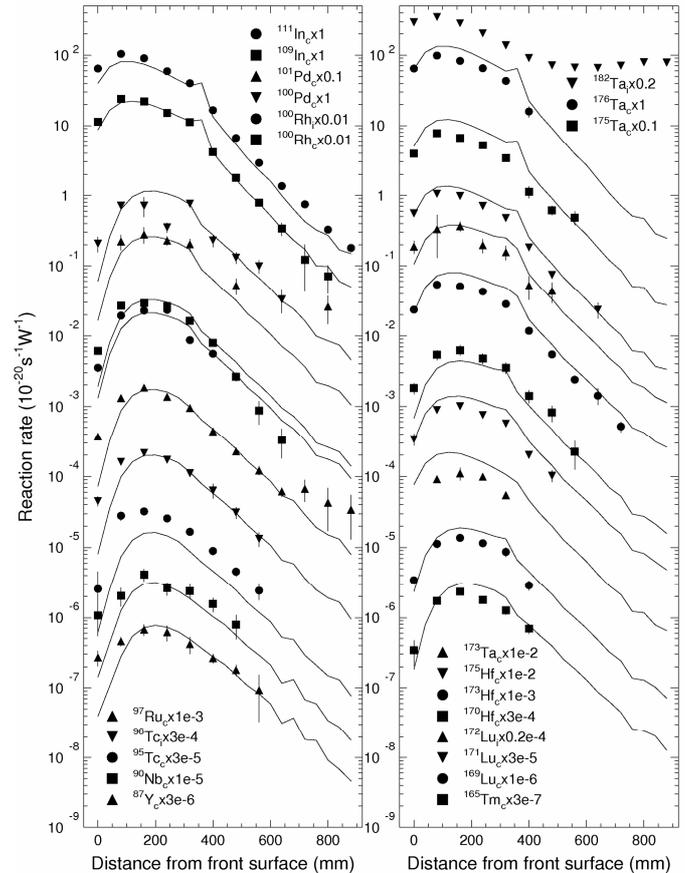

Fig. 5. Experimental and calculated distributions of the $^{nat}$In (left panel) and $^{181}$Ta (right panel) reaction products along the surface of the Pb target.